\documentclass[fleqn,usenatbib]{mnras}

\usepackage[T1]{fontenc}
\usepackage{ae,aecompl}

\usepackage{graphicx}	
\usepackage{amsmath}	
\usepackage{amssymb}	

\usepackage{cprotect}

\newcommand{\eff}{_{\textrm{\tiny eff}}} 

\title[{\it Gaia} DR2 ELM Catalogue]{{\it Gaia} Data Release 2 Catalogue of Extremely-low Mass White Dwarf Candidates}

\author[Pelisoli \& Vos]{
Ingrid Pelisoli\thanks{E-mail: pelisoli@astro.physik.uni-potsdam.de} \& Joris Vos
\\
Institut f\"{u}r Physik und Astronomie, Universit\"{a}t Potsdam, Haus 28, Karl-Liebknecht-Str. 24/25, D-14476 Potsdam-Golm, Germany
}

\date{Accepted XXX. Received YYY; in original form ZZZ}

\pubyear{2019}

\begin{document}
\label{firstpage}
\pagerange{\pageref{firstpage}--\pageref{lastpage}}
\maketitle

\begin{abstract}
Extremely-low mass white dwarf stars (ELMs) are $M < 0.3~M_{\odot}$ helium-core white dwarfs born either as a result of a common-envelope phase or after a stable Roche-lobe overflow episode in a multiple system. The Universe is not old enough for ELMs to have formed through single-star evolution channels. As remnants of binary evolution, ELMs can shed light onto the poorly understood phase of common-envelope evolution and provide constraints to the physics of mass accretion. Most known ELMs will merge in less than a Hubble time, providing an important contribution to the signal to be detected by upcoming space-based gravitational wave detectors. There are currently less than 150 known ELMs; most were selected by colour, focusing on hot objects, in a magnitude-limited survey of the Northern hemisphere only. Recent theoretical models have predicted a much larger spacial density for ELMs than estimated observationally based on this limited sample. In order to perform meaningful comparisons with theoretical models and test their predictions, a larger  well-defined sample is required. In this work, we present a catalogue of ELM candidates selected from the second data release of {\it Gaia} (DR2). We have used predictions from theoretical models and analysed the properties of the known sample to map the space spanned by ELMs in the {\it Gaia} Hertzsprung-Russell diagram. Defining a set of colour cuts and quality flags, we have obtained a final sample of  5762 ELM candidates down to $T\eff \approx 5000$~K.
\end{abstract}




\section{Introduction}

White dwarf stars are by far the most common outcome of the evolution of single stars, representing the final observable stage of all stars with initial mass lower than 7--11.8~$M_{\odot}$ \citep{woosley2015, lauffer2018}. Given the age of the Universe and the timescales of single stellar evolution, the minimal mass for white dwarfs formed through this channel is around 0.3~$M_{\odot}$ \citep{kilic2007}. Objects below this limit can only be explained as outcomes of the evolution of binary or multiple systems. Although binary fraction in the main sequence depends on initial mass, going from 0.72 for O stars \citep{mason1998} to $0.26\pm0.1$ for late-type M stars \citep{basri2006}, for stars that have had enough time to evolve off the main sequence ($M \gtrsim 0.85~M_{\odot}$), the binarity fraction seems to be larger than 0.5 \citep{duchene2013}. Although the exact fraction depends on metallicity \citep{moe2019}, around 25 per cent of these systems should be close enough to interact during their lifetime, by exchanging mass and angular momentum \citep{willems2004}. This interaction influences their evolution and gives rise to new astrophysical phenomena otherwise absent from the life of single stars. One such example of exotic objects resulting only from binary interactions are the extremely low-mass white dwarfs (ELMs).

ELMs show $M < 0.3~M_{\odot}$ and an observed binarity rate consistent with 100 per cent \citep{elmsurveyVII}. Given the inverse relation between mass and radius for white dwarfs, they show larger radius and are therefore brighter than canonical mass white dwarfs. Yet their existence was elusive for many years, because their low-mass implies surface gravity and colours outside the selection region of canonical white dwarfs. Moreover, as they are brighter than canonical white dwarfs, they do not necessarily stand out as high proper motion objects in magnitude-limited surveys. The catalogue of \citet{bergeron1992} did serendipitously include 14 objects with mass below 0.45~$M_{\odot}$, which is below the single evolution limit considering canonical mass-loss mechanisms and solar metallicity \citep{kilic2007}. Following-up seven of these objects, \citet{marsh1995} found five of them to be in close binaries, suggesting binarity was necessary for their formation. More recent results suggest a binarity rate of around 70 per cent, indicating that although the binary scenario is dominant, other channels might be required, such as mass loss due to winds or interactions with substellar companions \citep{brown2011}. Alternatively, the objects currently observed as single stars might have formed in a binary that was disrupted by a type Ia supernova or merged \citep{justham2009,justham2010}. These objects in the range 0.30--0.45~$M_{\odot}$ are commonly referred to as low-mass white dwarfs.

The first spectroscopically confirmed ELM was found as a companion to pulsar PSRJ1012+5307 \citep{kerkwijk1996}, following other optical identifications of pulsar companions that are likely faint ELMs \citep{bell1993,lorimer1995}. A similar field object was only found considerably later by \citet{heber2003}. Initially classified as a hot subdwarf star, it was found to have a mass of only $0.24~M_{\odot}$, therefore too low to sustain core helium burning, leading to a new classification as a He core white dwarf. Time-resolved spectroscopy confirmed the object to be in a binary with an unseen companion, with a minimum mass of $0.73~M_{\odot}$, which therefore must be a compact degenerate object. Both the initial classification of this object and the fact that all similar others had been found in binaries lead to a parallel between the formation of ELMs and of hot subdwarfs: both are formed by enhanced mass-loss due to either (i) common-envelope evolution or (ii) stable Roche-lobe overflow. In the case of hot subdwarfs, this happens after the onset of core-He burning \citep{han2002, han2003}, whereas for ELMs a degenerate He-core of a low-mass progenitor is exposed prior to the ignition of He in the core \citep{istrate2016, li2019}. A third scenario explains the existence of single hot subdwarfs, which correspond to around 60 per cent of the observed population \citep{heber2016}: a merger between two He-core white dwarfs \citep{zhang2012}. Although single ELMs could also result from mergers, or from the remnants of giant-branch donor stars whose envelopes have been stripped off by the supernova explosion \citep{justham2009}, no such object has yet been detected. However, given the difficulty of precisely obtaining atmospheric parameters from the Balmer lines at the range of effective temperature ($T\eff \lesssim 20\,000$~K) and surface gravity ($ 4.75 \lesssim \log~g \lesssim 7.0$) of the ELMs \citep{brown2017, pelisoli2018}, discoveries have actually relied on orbital parameters to constrain the radius of the observed candidates and confirm or rule out their nature as ELMs.

There are currently around 140 known ELMs as primaries in a binary system. More than half of this sample was discovered as part of the ELM Survey \citep{elmsurveyI, elmsurveyII, elmsurveyIII, elmsurveyIV, elmsurveyV, elmsurveyVI, elmsurveyVII}, which followed-up objects selected from their colours from the Sloan Digital Sky Survey \citep[SDSS, ][]{sdss}. Their survey mostly targeted objects hotter than 8500 K, introducing a temperature selection. Moreover, as it followed SDSS observations, the known sample contains very few southern objects and is magnitude-limited. Evolutionary models predict that, within the age of the Universe, ELMs could have cooled down to temperatures of $\approx 5000$~K \citep{istrate2016}. Therefore this selection for hot stars in the known population prevents a complete test of the models and hampers our progress in the binary evolution field. As remnants of binary evolution, ELMs can shed light onto the poorly understood yet crucial phase of common-envelope evolution \citep{nelemans2000} and are ideal laboratories for studying tidal effects \citep{fuller2013}. Additionally, most known ELMs will merge in less than a Hubble time resulting in new exotic objects such as extreme helium stars, R Corona Borealis stars, He-rich subdwarfs, and underluminous supernovae \citep{webbink1984, brown2016}. As strong sources of gravitational waves, they will have an important contribution to the signal detected by space-based missions such as LISA \citep{brown2016} allowing further development of the new field of multi-messenger astronomy \citep[e.g][]{tauris2018}.

In order to deliver the full diagnostic potential of ELMs, a sample covering the full range of physical parameters predicted by the models and as complete as possible is required. In this work, we take advantage of the groundbreaking {\it Gaia} data release 2 to select ELM candidates with well-defined selection criteria, covering the whole sky and the full-range of temperatures predicted by the models, taking the first step towards an all-sky, complete, and unbiased ELM sample. In Section~\ref{sec:knownELMs}, we describe the known population of ELMs, which we combine with predictions from evolutionary models to map the selection region in the {\it Gaia} DR2 HR diagram. In Section~\ref{sec:selecion}, we describe in details our selection procedure. Section~\ref{sec:catalogue} presents the resulting catalogue and its properties, and Section~\ref{sec:end} contains summary and conclusions.

\section{The known ELM sample}
\label{sec:knownELMs}

In order to map the parameter space occupied by ELMs in the {\it Gaia} observational HR diagram, absolute magnitude $G_\textrm{abs}$ as a function of the colour $G_{BP} - G_{RP}$, we have first compiled a catalogue of known ELMs from the literature. It updates table~4 of \citet{silvotti2012}, including all the objects from table~5 of \citet{elmsurveyVII}, table~1 of \citet{pelisoli2017}, and table~3 of \citet{pelisoli2018b}, as well as other objects individually published in the literature. We have not included at this point ELMs found as companions to millisecond-pulsars, because they are mostly too faint to be detectable by {\it Gaia}, or even to be fully confirmed ELMs, but a list of such objects can be found in Appendix~\ref{sec:apA}. We have also refrained from including systems in which \mbox{(pre-)ELMs} are the secondary star, the so-called EL~CVn binaries \citep[named after their prototype, e.g][]{maxted2014,vanroestel2018}. These systems are dominated by the primary main sequence star, and therefore overlap with the main sequence in the {\it Gaia} HR diagram. The ELMs in these systems have only been discovered because they eclipse the primary star. In short, our list only includes systems in which ELMs are primaries, whose companions are canonical-mass white dwarfs or low-mass main sequence stars. Moreover, we do not make a difference between pre-ELMs and ELMs, given that the complicated evolution of these systems causes the evolutionary tracks of the pre-white dwarf and white dwarf phase to overlap \citep[see e.g. fig.~8 of][]{li2019}, therefore the differentiation between these two phases is not straightforward, and confusion exists in the literature.

\begin{figure}
	\includegraphics[width=\columnwidth]{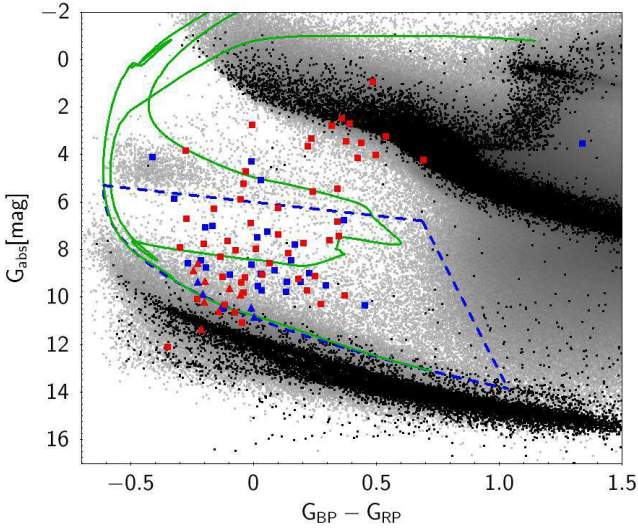}
    \cprotect\caption{Colour-magnitude diagram showing known ELMs from Table~\ref{tab:elm_cat} in blue (when the value of parallax over its uncertainty, $\verb!parallax_over_error!$, is larger than 5) or red ($\verb!parallax_over_error! < 5$), as squares for $M < 0.3~M_{\odot}$, and triangles for $M > 0.3~M_{\odot}$. We show in black the Gaia 100~pc clean sample \citep[Sample C in][]{lindegren2018}, and in grey a comparison sample of objects following the same quality criteria of \citet{lindegren2018}, but with $\verb!parallax! > 1$ and $\verb!parallax_over_error! > 5$. These two comparison samples will be shown throughout the paper following the same colour code. The green solid line shows an evolutionary model by \citet{istrate2016}, mapped onto the {\it Gaia} passbands. The final mass for this model is $0.324~M_{\odot}$, the highest published in \citet{istrate2016}. We defined a colour-cut delimited by the blue dashed lines to contain most of the known ELMs with $\verb!parallax_over_error! > 5$ and $M < 0.3~M_{\odot}$, using the evolutionary model as a bottom limit in order to exclude higher mass WDs (below the selection), and defining further cuts to exclude hot subdwarfs (above), and scattered main sequence stars (to the right).}
    \label{fig:known_elms}
\end{figure}

The resulting catalogue of known ELMs is shown in Table~\ref{tab:elm_cat}, and contains 119 objects. We have cross-matched this list with {\it Gaia} DR2, obtaining one match for each star, as indicated by the {\it Gaia} DR2 identifier $\verb!source_id!$ in Table~\ref{tab:elm_cat}. Six objects have only a two-parameter (right ascension $\alpha$ and declination $\delta$) solution, whereas the remaining 113 stars have the complete five-parameter solution ($\alpha$, $\delta$, parallax $\pi$, and the proper motions $\mu_{\alpha}$ and $\mu_{\delta}$). Figure~\ref{fig:known_elms} shows the location of these objects in the {\it Gaia} HR diagram.

\begin{table*}
	\centering
	\caption{Compiled catalogue of known ELMs. The complete table can be found in the online version of the paper.}
	\label{tab:elm_cat}
\begin{tabular}{cccccccccccc}
\hline
\hline
  ID & RA & DEC & Gaia DR2 & $T\eff$ & $\log~g$ & $M_1$ & $M_2$ (min) & Reference \\
  	 & (J2000) & (J2000) & source\_id & (K) & [cgs] & $(M_{\odot})$ &  $(M_{\odot})$ \\ 
\hline
  J0022-1014 & 00:22:07.66 & -10:14:23.5 & 2425129334949091840 & 20730 & 7.28 & 0.376 & 0.21 & \citet{elmsurveyVII} \\
  J0022+0031 & 00:22:28.44 & +00:31:15.6 & 2546819845937777536 & 20460 & 7.58 & 0.459 & 0.23 & \citet{elmsurveyVII} \\
  J0042-0106 & 00:42:27.73 & -01:06:34.8 & 2530847858996083456 & 8051 & 5.51 & 0.145 & 0.14 & \citet{pelisoli2018b} \\
  J0056-0611 & 00:56:48.22 & -06:11:41.4 & 2524390049249274112 & 12230 & 6.17 & 0.18 & 0.46 & \citet{elmsurveyVII} \\
  J0106-1000 & 01:06:57.38 & -10:00:03.1 & 2470207353182651008 & 16970 & 6.1 & 0.189 & 0.39 & \citet{elmsurveyVII} \\
  J0112+1835 & 01:12:10.24 & +18:35:04.1 & 2786627626922933248 & 10020 & 5.76 & 0.16 & 0.62 & \citet{elmsurveyVII} \\
  J0115+0053 & 01:15:08.65 & +00:53:46.2 & 2535267895739445120 & 8673 & 5.64 & 0.15 & 0.05 & \citet{pelisoli2018b} \\
  J0125+2017 & 01:25:16.76 & +20:17:44.7 & 2786529117553328640 & 11170 & 4.71 & 0.184 & 0.14 & \citet{elmsurveyVII} \\
  J0152+0749 & 01:52:13.78 & +07:49:14.3 & 2568823856748296832 & 10840 & 5.92 & 0.169 & 0.57 & \citet{elmsurveyVII} \\
  J0306-0013 & 03:06:08.90 & -00:13:38.9 & 3266357782915336320 & 7768 & 5.36 & 0.143 & 1.03 & \citet{pelisoli2018b} \\
\hline
\hline
\end{tabular}
\end{table*}

Most of the catalogued ELMs are placed between the main sequence and canonical mass white dwarfs, as expected given their intermediary radii. About ten objects are overlapping the hot subdwarf region, with two of them found on top of the hot subdwarf cloud \citep[HS2231+2441 and HE0430-2457 from][respectivelly]{almeida2017,vos2018}. In both these cases, the objects were initially classified as hot subdwarfs, but their masses have been found to be lower than the required for core-He burning, leading to the classification as pre-ELMs. There is also a considerable number of objects lying closely to the main sequence, which could also be better classified as pre-ELMs. Before reaching the cooling sequence, ELMs can show radii of the same order as main sequence or even horizontal branch stars, due to residual shell burning. Finally, there is one star (J0308+5140) which stands out, lying much redder than the remaining of the sample. As pointed out in its discovery paper \citep{elmsurveyV}, this object is in a direction with considerable reddening. As in \citet{fusillo2019}, we apply no reddening correction during our selection. The main reason for this choice is the poor resolution in reddening maps. In {\it Gaia} DR2 extinction and reddening are provided for many stars, but they caution that these values should not be applied to individual stars; they are adequate for population studies where averages over a property are the variable of interest \citep[e.g][]{andrae2018}.

A few published ELMs in Fig.~\ref{fig:known_elms} lie closely to the canonical mass white dwarf cooling sequence; their published masses are between $0.35$ and $0.49~M_{\odot}$. This would suggest that they are better classified as low-mass white dwarfs \citep{kilic2007}. Curiously, they lie above the evolutionary model of \citet{istrate2016} shown in Fig.~\ref{fig:known_elms}, which would therefore suggest a mass lower than $0.324~M_{\odot}$. The reddening for these objects is not significant \citep[$E(B-V) = 0.1321$ for one object, but smaller than 0.05 for the remaining others according to][]{schlafly2011}, and applying a correction does not move them below the model. We do not know the reason for this discrepancy, but it likely reflects some inconsistency between the evolutionary models and spectral modes used to determine the $\log~g$ of observed ELMs. We will further discuss how this might impact our selection in Section~\ref{sec:catalogue}; for now we rely on the limit defined by the evolutionary models, since our goal is to obtain a benchmark sample to test such models.

\section{ELM Candidate Selection}
\label{sec:selecion}

\subsection{{\it Gaia} DR2 data}
\label{sec:gaia}

Analysing the position of the known ELMs with $M < 0.3~M_{\odot}$ and $\verb!parallax_over_error! > 5$, as well as the position an evolutionary model corresponding to the upper mass limit in Fig.~\ref{fig:known_elms}, we have defined an initial colour-cut to select ELM candidates, illustrated also in Fig.~\ref{fig:known_elms}. We select objects in the region delimited by the three following functions:
\begin{eqnarray}
G_\textrm{abs} &=& 5.25 + 6.94\,(G_{BP} - G_{RP}+0.61)^{1/2.32}; \label{sel1_bottom}\\
G_\textrm{abs} &=& 1.15\, (G_{BP} - G_{RP}) + 6.00; \label{sel1_top} \\
G_\textrm{abs} &=& 20.25\, (G_{BP} - G_{RP}) - 7.15; \label{sel1_right}
\end{eqnarray}
where $G_\textrm{abs} = G + 5 \log (( \verb!parallax!+0.029)/1000) + 5$ \citep[we have assumed the parallax zero-point given by][determined using quasars, which have similar colours to ELMs]{lindegren2018}. The line at the bottom of the selection, Eq.~\ref{sel1_bottom}, aims at excluding canonical mass white dwarfs from the selection, and was defined to fit the higher-mass evolutionary model of \citet{istrate2016}, with final mass of $0.324~M_{\odot}$. A similar criterion with a low mass model cannot be applied, because ELMs can experience flashes during their evolution \citep[see e.g. fig.~2 in][]{corsico2014}, leading to luminosities as high as main sequence stars. Therefore the low mass model cannot be used to place an upper limit in magnitude. Instead the top line, Eq.~\ref{sel1_top}, was defined to keep hot subdwarfs out of the selection. Finally, the line at the right, Eq.~\ref{sel1_right}, aims at excluding stars scattered from the main sequence due to statistical uncertainties. We placed this line as red as possible to include cool ELMs in our selection, but we further refine its position in Sec.~\ref{sec:sdss}. Whereas this selection might be excluding objects still in the pre-ELM phase, it should include the bulk of objects already in the cooling track.

We also apply quality filter parameters to the selection, following section~2.1 of \citet{pelisoli2019}. This selection is very similar to the one in Appendix~B of \citet{lindegren2018}, but slightly relaxes the requirements on the Gaia DR2 value of $\verb!phot_bp_rp_excess_noise!$, $E$, based on observed values for the known ELMs. The criteria are:
\begin{eqnarray}
1.0 + 0.015\,(G_{BP} - G_{RP})^2 &<& E\\
E &<& 1.45 + 0.06\, (G_{BP} - G_{RP})^2 \nonumber
\end{eqnarray}
and
\begin{eqnarray}
u &<& 1.2\, \verb!max!(1, \exp(-0.2(G - 19.5)))
\end{eqnarray}
where $$u = \sqrt{ \verb!astrometric_chi2_al!/(\verb!astrometric_n_good_obs_al! - 5)}$$
$\verb!astrometric_chi2_al!$ and $\verb!astrometric_n_good_obs_al!$ are, respectively, the value of the chi-square statistic of the astrometric solution and the number of good observations; both are given in the {\it Gaia} DR2 table. We initially only restrict the parallax uncertainty to $\verb!parallax_over_error! > 3$. The ADQL query including all of these criteria is given in Appendix~\ref{ref:initial}, and results on 123\,945 objects.

\subsection{SDSS spectra}
\label{sec:sdss}

\begin{figure}
	\includegraphics[width=\columnwidth]{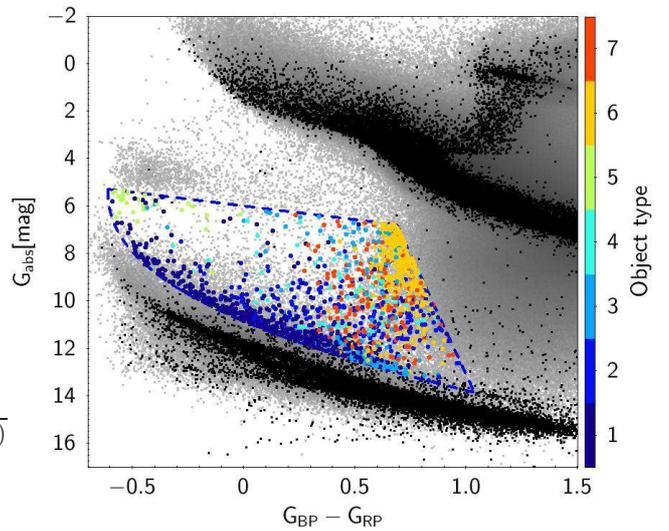}
    \caption{Colour-magnitude diagram showing objects with SDSS spectra colour-coded by their spectral type, where 1 = WD, 2 = WD+MS, 3 = A, 4 = CV, 5 = B, 6 = F, and 7 = QSO. For a description of what each of these spectral classifications entails, see the text. Note the large contamination of F stars and QSOs given our initial selection criteria.}
    \label{fig:selection2}
\end{figure}

\begin{figure}
	\includegraphics[width=\columnwidth]{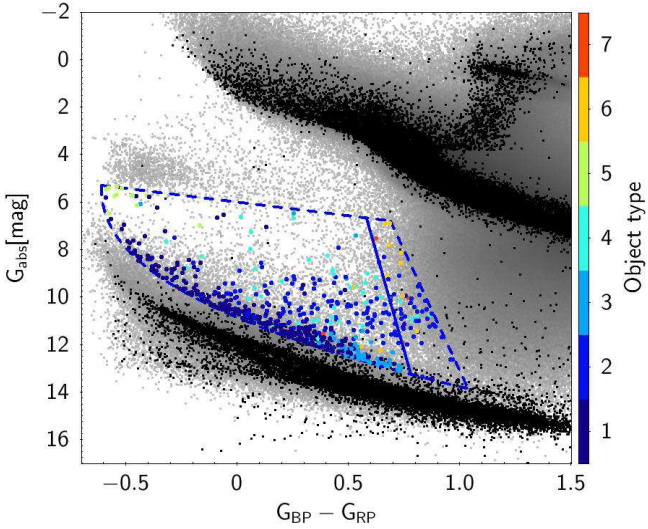}
    \cprotect\caption{Same as Fig.~\ref{fig:selection2}, but showing only SDSS objects where $\verb!parallax_over_error! > 5$. The contamination by F stars is reduced to 14 objects, and only four quasars remain in the sample. In order to exclude most of these contaminants, we redefine the red border of our selection to the solid blue line.}
    \label{fig:selection3}
\end{figure}

We next cross-match the result of our initial selection with the Sloan Digital Sky Survey (SDSS) spectral database, retrieving spectra for 1\,492 of the selected objects. These spectra were visually inspected and attributed general classifications: $B$ (H + He lines), $A$ (H lines), $F$ (H lines + metals), $CV$ (cataclysmic variable, H or He in emission), $QSO$ (quasar, broad and red-shifted emission lines), $WD$ (canonical white dwarf, broad absorption lines), or $WD+MS$ (canonical white dwarf with cool main sequence companion, composed spectra). The WD classification includes objects of $A$ and $B$ type, but with obviously broad absorption features. Fig.~\ref{fig:selection2} illustrates the result of our visual inspection. It is clear that there is a large contribution of scattered F stars (24 per cent of the sample with spectra) and even QSOs (11 per cent) to the selected region. We verified, however, that increasing our restriction in parallax uncertainty to $\verb!parallax_over_error! > 5$ eliminates almost fully this contamination (see Fig.~\ref{fig:selection3}). With this criterion, there remain only four quasars and 14 F stars (0.7 and 2.3 per cent of the sample with spectra and $\verb!parallax_over_error! > 5$, respectively). One QSO is well-within our selection, close to the white dwarf cooling track. The other three, however, are close to the red border of our selection. We modify this border in order to exclude these objects, in the hopes of limiting further contamination by QSOs in the selected sample. The new red limit of our selection is given by
\begin{eqnarray}
G_\textrm{abs} &=& 33\,(G_{BP} - G_{RP})- 12.5.
\end{eqnarray}
This also reduces the number of identified F stars in the sample to four. With this new limit and $\verb!parallax_over_error! > 5$, our query (given in Appendix~\ref{sec:refined}) now results on 11\,993 objects, which we will refer to as ``refined sample". Given that our selection extends to the very limit of the evolutionary model, it is selecting ELMs with temperatures down to $\approx 5000$~K.

\subsection{Further colour cuts}
\label{sec:red_border}

\begin{figure}
	\includegraphics[width=\columnwidth]{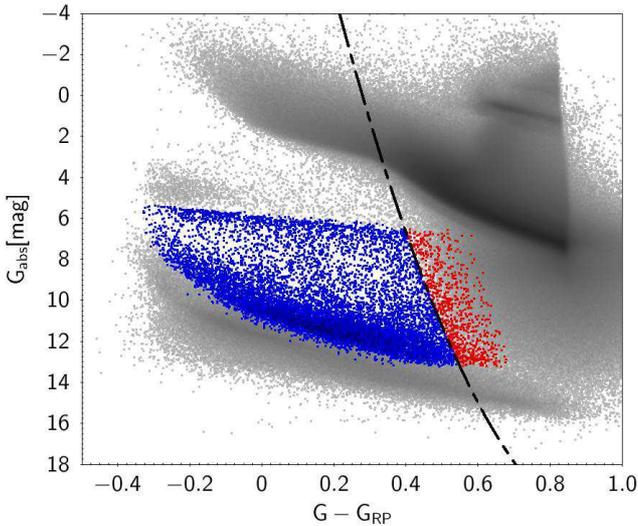}
    \caption{Absolute magnitude $G_\textrm{abs}$ as a function of colour $G - G_{RP}$. This diagram is suggested to be more adequate to analyse faint red stars; we use it to further refine the red limit of our sample, excluding objects redder than the dot-dashed line, which overlap with a large cloud of scattered main sequence stars.}
    \label{fig:selection4}
\end{figure}

\citet{gaia2018} cautions that faint sources might show a turn towards the blue at the lower end of the main sequence, resulting from a stronger flux excess in $G_{BP}$ than in $G_{RP}$, suggesting that the $G - G_{RP}$ colour might be more adequate to study faint red sources. Given our concern with faint red stars scattered from the main sequence into our selected region, we analysed the $G_\textrm{abs} \times G - G_{RP}$ diagram for our selected objects, as shown in Fig.~\ref{fig:selection4}. A cloud of scattered stars is clearly visible in this diagram. To reduce contamination from this objects, we apply one further selection to our sample
\begin{eqnarray}
G_\textrm{abs} > -42.2\,(G-G_{RP})^2 + 83.8\,(G-G_{RP}) - 20.1
\end{eqnarray}
11\,167 objects are left following this cut.

Another issue to be considered are close visual neighbours, which have been shown to affect the {\it Gaia} measurements without that being apparent in any quality control flags \citep[e.g][]{fusillo2019,boubert2019}. To avoid that, we have removed from our sample any objects with another {\it Gaia} source within 5~arcsec. There remain 8009 objects with no detected neighbours within 5~arcsec.

\begin{figure}
	\includegraphics[angle=-90,width=\columnwidth]{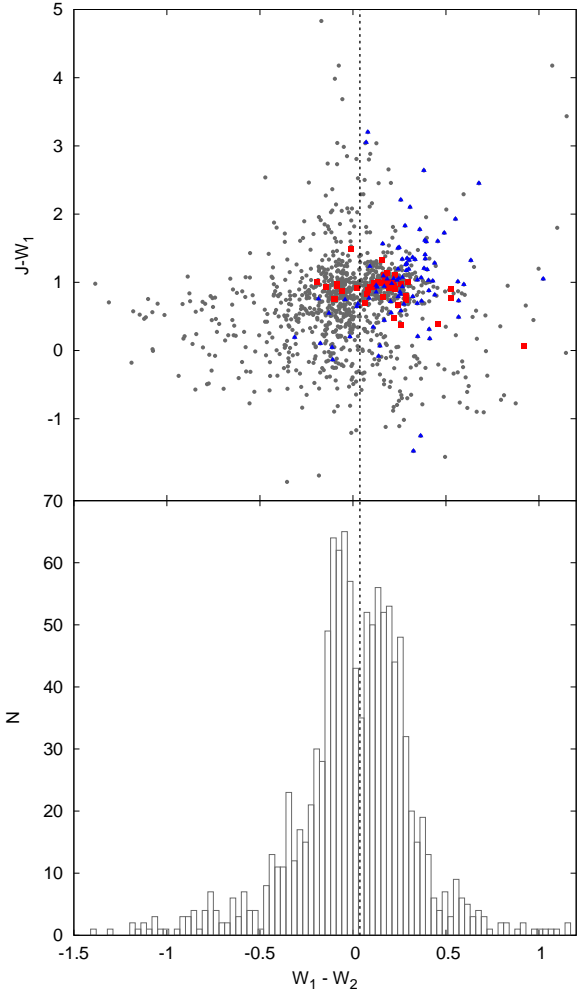}
    \caption{Colour-colour diagram showing $(J-W_1)\times(W_1-W_2)$ (top), and a histogram of $(W_1-W_2)$ (bottom). In both panels, it is clear that CVs (blue triangles) and WD+M (red squares) concentrate redder than $(W_1 - W_2 = 0.0375)$.}
    \label{fig:selection5}
\end{figure}

These selection criteria and those implemented in Section~\ref{sec:sdss} are aimed at efficiently removing scattered main sequence stars and quasars from the selected region of the HR diagram; however, such region is also populated by CVs and canonical mass white dwarfs with main sequence companions. In order to separate this type of objects from the ELM candidates, we have relied on further colour cuts based on colour-colour diagrams. We cross-matched our sample with 2MASS \citep{2mass} and WISE \citep{wise} (all sky surveys), SDSS and PanSTARRS \citep{panstarrs} (Northern sky), and SkyMapper \citep{skymap} (Southern sky). Fig.~\ref{fig:selection5} shows a $(J-W_1)\times(W_1-W_2)$ diagram. It is clear that objects identified as CVs or white dwarfs with main sequence companions from the SDSS spectra are concentrated in a region of the diagram, leading to the colour cut
\begin{eqnarray}
W_1 - W_2 < 0.0375. \label{eq:wise}
\end{eqnarray}

\begin{figure}
	\includegraphics[angle=-90,width=\columnwidth]{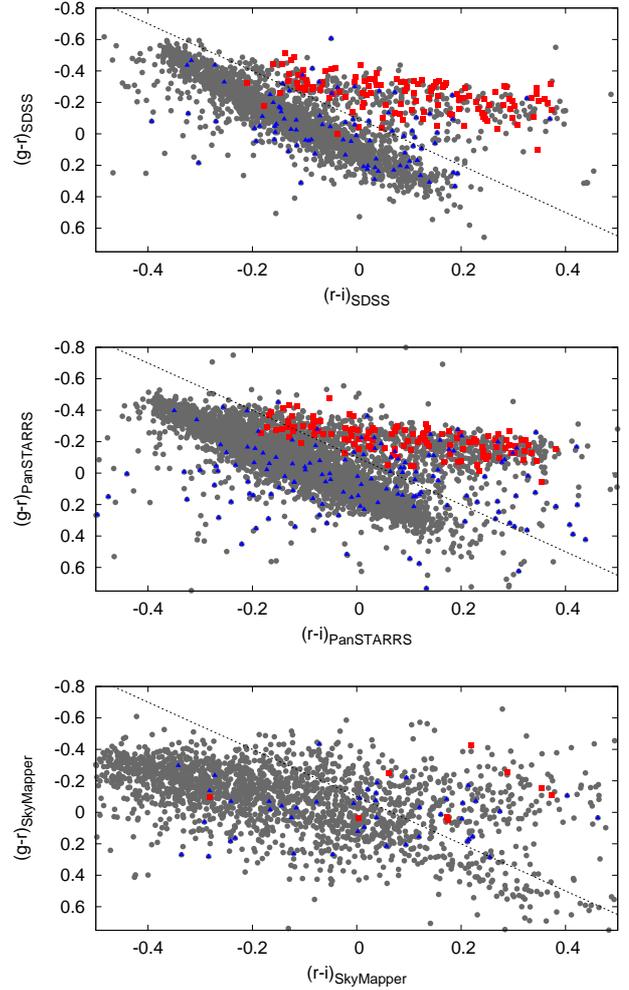}
    \caption{Colour-colour diagrams for $(g-r)\times(r-i)$. The top diagram uses SDSS colours, the middle uses PanSTARRS, and the bottom SkyMapper. The colour code is the same as in Fig.~\ref{fig:selection5}. Most WD+M and a few CVs concentrate redder than the dashed lines, which were used as further selection criteria.}
    \label{fig:gri}
\end{figure}

\begin{figure}
	\includegraphics[angle=-90,width=\columnwidth]{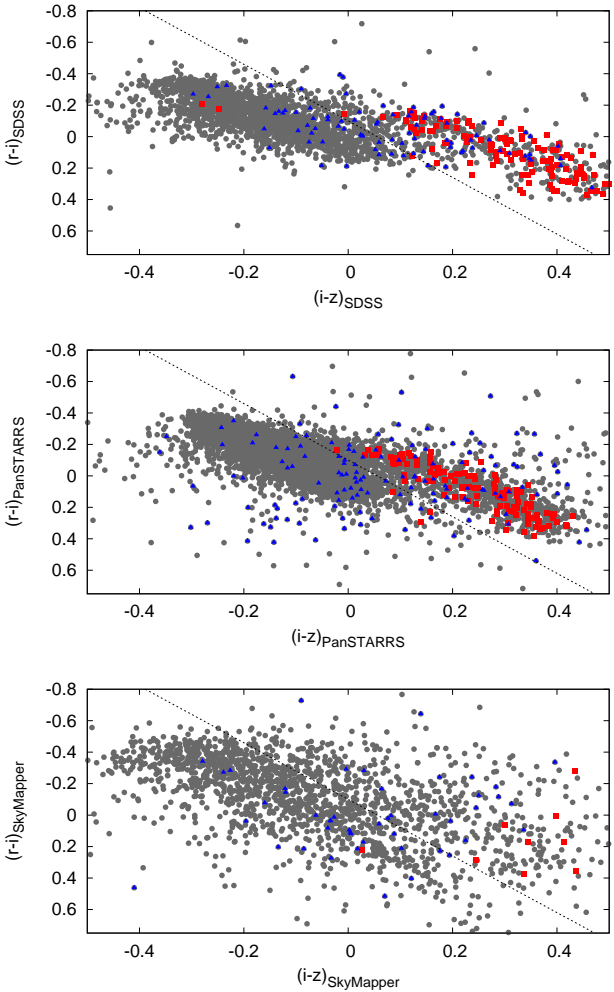}
    \caption{Same as Fig.~\ref{fig:gri}, but for the colours $(r-i)\times(i-z)$.}
    \label{fig:riz}
\end{figure}

\begin{figure}
	\includegraphics[angle=-90,width=\columnwidth]{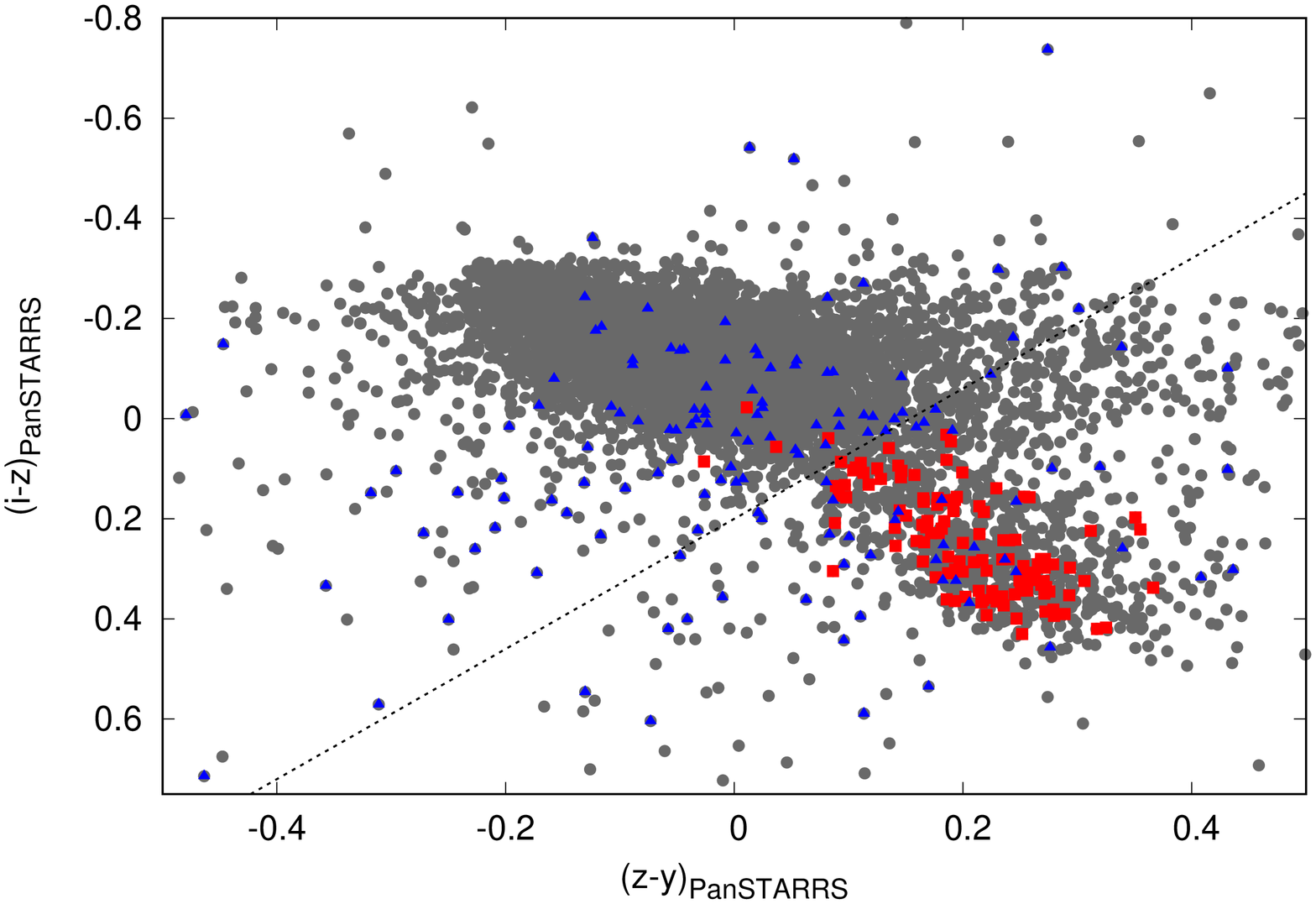}
    \caption{Colour-colour diagram for $(i-z)\times(z-y)$, using PanSTARRS, with the same colour code as Fig.~\ref{fig:riz}.}
    \label{fig:izy}
\end{figure}

In SDSS and PanSTARRS colour-colour diagrams, we can also see that these objects present a clear location in the $(g-r)\times(r-i)$ (Fig.~\ref{fig:gri}), $(r-i)\times(i-z)$ (Fig.~\ref{fig:riz}), and $(i-z)\times(z-y)$ (Fig.~\ref{fig:izy}) diagrams. We apply the following cuts:
\begin{eqnarray}
(g-r) > 1.5\,(r-i) - 0.1\\
(r-i) > 1.8\,(i-z) - 0.1\\
(i-z) < -1.3\,(z-y) + 0.2 \label{eq:panstarrs}
\end{eqnarray}
As these surveys cover only the Northern hemisphere, we rely on SkyMapper to a similar selection to Southern objects, although the location of objects with red-excess is not as clear for SkyMapper data, partially because we have much less objects with spectra in this region. We also exclude from the sample all objects identified from their spectra as CVs, WD+MS, QSO, or F star, or with SIMBAD classification as such. 5768 objects remain.

\begin{figure}
	\includegraphics[width=\columnwidth]{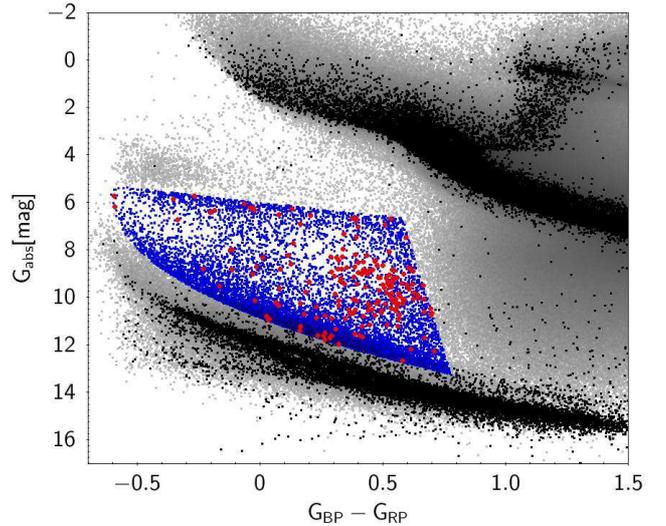}
    \caption{{Gaia} observational HR diagram showing the refined sample in blue dots, and objects with X-Ray detection as red diamonds.}
    \label{fig:selection6}
\end{figure}

To identify further CV and QSO candidates, we cross-match our sample with the Master X-Ray catalogue compiled by HEASARC\footnote{https://heasarc.nasa.gov/W3Browse/all/xray.html}. Fig.~\ref{fig:selection6} shows the objects within the refined sample with X-Ray measurements. They mostly correlate with known QSOs or CVs, but there are further six objects with known X-Ray emission but no classification. We separate those as well, being left with 5762 ELM candidates, which we will refer to as ``clean sample''.

The objects separated from the sample using the colour cuts from Eqs.~\ref{eq:wise}--\ref{eq:panstarrs} and the X-Ray flux are listed in Appendix~\ref{sec:redstars}. As candidate CVs and binary white dwarfs with main sequence companions, they remain of interest. It is important to mention that one ELM has been found with a main sequence companion \citep[J1555+2444, mentioned as exception by][]{elmsurveyVII}; therefore, by separating these objects from our ELM candidate sample, we might be excluding some ELMs. However, by doing so with a clear set of criteria, we maintain a sample easy to be simulated with population synthesis models in order to test the predictions of evolutionary models. In \citet{li2019}, for example, they predict a density of 1500~kpc$^{-3}$ for ELMs in double degenerate binaries, i.e. also excluding ELMs with main sequence companions. This prediction should be straightforward to test once our ELM candidates are confirmed.

\section{Catalogue of ELM Candidates}
\label{sec:catalogue}

\begin{figure}
	\includegraphics[width=\columnwidth]{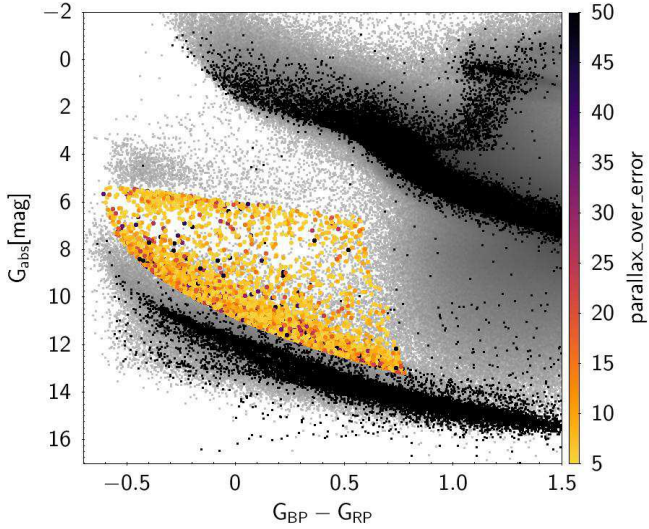}
    \cprotect\caption{{\it Gaia} observational HR diagram showing the selected clean sample of ELM candidates colour-coded by $\verb!parallax_over_error!$. The comparison samples are the same as in previous figures.}
    \label{fig:selection7}
\end{figure}

\begin{figure}
	\includegraphics[width=\columnwidth]{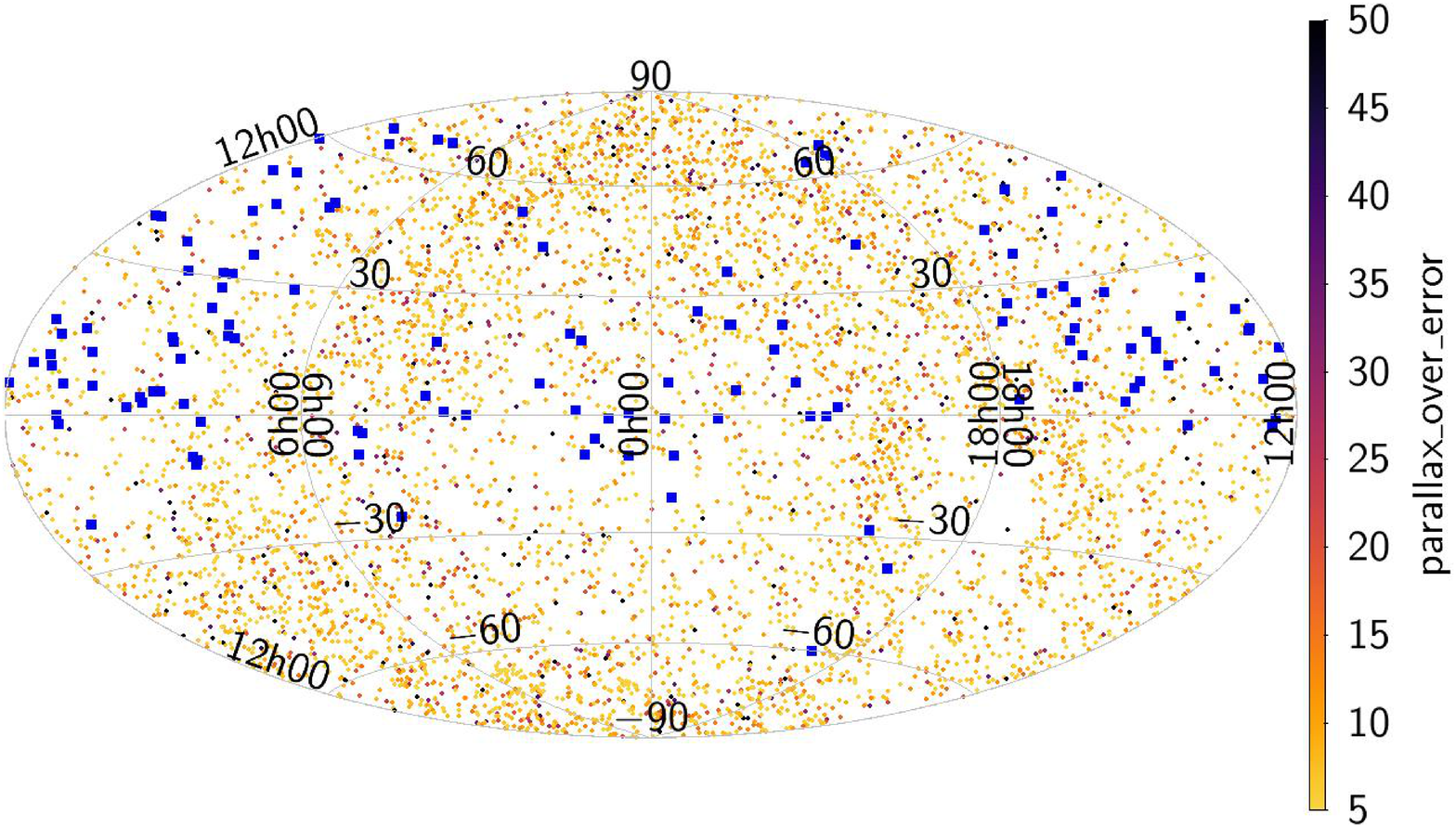}
    \cprotect\caption{Sky distribution of the selected clean sample of ELM candidates colour-coded by $\verb!parallax_over_error!$. The knwon ELMs from Table~\ref{tab:elm_cat} are shown as blue squares, and concentrate in the North hemisphere only.}
    \label{fig:sky}
\end{figure}

Our clean ELM candidate sample contains $5\,672$ objects, listed in Table~\ref{tab:elm_can_cat}. The HR diagram with this clean sample is shown in Fig.~\ref{fig:selection7}. The candidates are distributed over the whole sky, as can be seen in Fig.~\ref{fig:sky}. This is the first all-sky sample of candidate ELMs, given that previous searches relied on SDSS data. We extend as further in colour, and hence in temperature, as predicted by the models of \citet{istrate2016}, which suggest ELMs can cool down to around $5\,000$~K within the age of the Universe. Therefore, we also remove the temperature selection introduced in previous surveys. Our catalogue does not include ELMs with early-type main sequence companions; these objects should lie close to the main sequence in the HR diagram, since the luminosity is dominated by the main sequence star, and therefore are not within our colour selection. ELMs with cool main sequence companions are also likely excluded, since objects with detectable red excess were separated to avoid contamination by CVs or canonical white dwarfs with main sequence companions, which populate the same region of the observational HR diagram. These objects are listed separately in Appendix~\ref{sec:redstars}.

\begin{table*}
	\centering
	\caption{Catalogue of ELM candidates from {\it Gaia} DR2. The complete table can be found in the online version of the paper, including further columns.}
	\label{tab:elm_can_cat}
	\begin{tabular}{cccccccccccc}
	\hline
	\hline
	$\verb!source_id!$ & RA & DEC & Parallax & $\mu$ & G \\
  	 & (J2000) & (J2000) & (mas) & (mas/yr) & (mag) \\
  	 \hline
 2873402974373195008 & 00:00:07.82 & +30:46:06.4 & 3.825 & 53.244 & 19.540\\
 387152784468510848 & 00:00:53.37 & +46:53:37.1 & 0.746 & 5.606 & 16.945\\
 2447719110579100032 & 00:01:00.41 & -04:27:42.8 & 2.412 & 14.524 & 18.605\\
 4918290711348007424 & 00:01:16.96 & -58:59:58.3 & 1.475 & 8.839 & 16.258\\
 384460943781889024 & 00:02:22.37 & +42:42:13.5 & 2.261 & 22.201 & 18.054\\
 2333842347693338496 & 00:02:31.28 & -26:52:19.5 & 3.491 & 44.514 & 19.215\\
 4684032857038291712 & 00:02:45.72 & -77:17:38.8 & 1.352 & 19.530 & 19.139\\
 2853174503041704832 & 00:02:59.59 & +25:26:08.5 & 1.516 & 21.146 & 19.027\\
 4688343801614700672 & 00:03:31.18 & -74:27:09.3 & 2.207 & 28.546 & 19.698\\
 2798386113507604992 & 00:03:33.93 & +20:16:26.4 & 1.736 & 17.231 & 18.984\\
 \hline
	\hline
	\end{tabular}
\end{table*}

Previous to the removal of objects classified as CV, QSO, WD+M, or F from our sample, there are 7 CVs, 3 F stars, and 15 objects classified as planetary nebulae (PN) in the sample, out of 274 objects with attributed classification (either from SIMBAD as different types of CV or PN, or from SDSS spectra), suggesting a contamination of the order of 9 per cent in our clean sample by these type of objects. However, given the apparent discrepancy between the evolutionary model used to determine the colour cuts and the mass obtained by spectral fits, there could also be contamination by low-mass ($0.30-0.45~M_{\odot}$) or canonical mass white dwarfs in the sample, as we further discuss below.

\begin{figure}
	\includegraphics[width=\columnwidth]{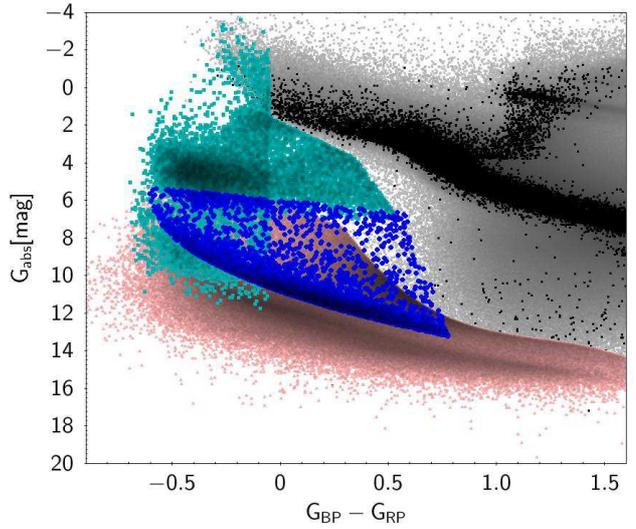}
    \cprotect\caption{Comparison between our clean sample (blue dots), the white dwarf catalogue of \citet{fusillo2019} (salmon triangles), and the hot subdwarf catalogue of \citet{geier2019} (turquoise squares).}
    \label{fig:comp}
\end{figure}

Our catalogue has overlap with the white dwarf catalogue of \citet{fusillo2019} and with the hot subdwarf catalogue of \citet{geier2019} (see Fig.~\ref{fig:comp}). 5368 of our candidates are also listed in the white dwarf catalogue, 4009 of them with calculated probability of being white dwarfs $P_{WD} > 0.75$, 3927 with mass estimates. 37 per cent of the masses are estimated to be smaller than $0.3~M_{\odot}$, therefore consistent with ELM. Further 58 per cent show $0.3 < M < 0.45~M_{\odot}$. Only 5 per cent show masses above $0.45~M_{\odot}$, i.e. that can be explained by single evolution without invoking mass-loss enhancement. This highlights the potential of our sample for testing binary evolution models. It is important to keep in mind, however, that masses estimated from photometry alone are highly uncertain, therefore spectroscopy needs to be obtained to reliably estimate the masses of these stars. The fact that a substantial number of the objects of our catalogue have $P_{WD} < 0.75$ reflects the fact that the catalogue of \citet{fusillo2019} focuses on canonical white dwarf stars, and although it does include ELMs, it is not in any way complete for them, as the authors themselves point out. 272 of our ELM candidates are in the hot subdwarf catalogue, and 360 objects are not in any of these catalogues.

As in other {\it Gaia} catalogues \citep[e.g][]{fusillo2019,geier2019}, we have not taken reddening into account for our colour selection. Values of $E(B-V)$ and $A_V$ from \citet{schlafly2011} are included in the extended online version of Table~\ref{tab:elm_can_cat}. About 20 per cent of the sample is in directions with significant extinction ($A_V > 1.0$~mag); however, taking into account the poor resolution (typically of several degrees) of dust maps, as well as the fact that total Galactic extinction is likely inadequate for nearby sources, we opt for not excluding any objects based on extinction or reddening.

Our main limitation is the constraint on $\verb!parallax_over_error! > 5$, which should exclude objects in distant or dense regions. The completeness of our catalogue is strongly affected by this choice. In the full {\it Gaia} DR2 catalogue, this is the median $\verb!parallax_over_error!$ value for $G = 16.5$. For the known ELMs of Table~\ref{tab:elm_cat}, the average absolute G magnitude is 8. Assuming as limiting magnitude the value of 16.5, the distance modulus is 8.5, corresponding to a distance of 500~pc, within which our catalogue of candidates should be fairly complete, except for dense regions such as close to the Galactic plane, where the {\it Gaia} coverage is worse and because we have removed objects with close visual companions.

We use this magnitude-limited sample ($G < 16.5$), excluding the disc region ($b < 30^{\circ}$), where the {\it Gaia} completeness drops and reddening is significant, to estimate the local density of ELMs, assuming the {\it Gaia} saturation limit to be $G=3$. Following the same approach as \citet{brown2016}, we use the $1/V_{max}$ method \citep{schmidt1975}. To take into account the shape of the Galaxy, we adopt the density model of \citet{juric2008}. As this model considers distinct density functions for the three Galactic components (halo, thick disc, and thin disc), it is necessary to separate the sample into these components. As we lack information on radial velocities, we rely solely on the tangential velocity ($v_T$) derived from {\it Gaia} proper motions and parallax. Based on the criteria of \citet{babusiaux2018}, we assume objects with $v_T < 40$~km/s belong to the thin disk, and objects with  $v_T > 200$~km/s are halo objects (there actually none with $G < 16.5$), with remaining objects belonging to the thick disc. With these assumptions and taking into account the 9~per cent contamination, we derive a local ELM space density of 275~kpc$^{-3}$. Using the same Galactic model, \citet{brown2016} estimate a density of 160~kpc$^{-3}$. However, they caution that their estimates are not the local space density of all ELM WDs, but describe a clean sample of binaries relevant for their merger rate estimates, implying that the total number of ELMs can only be larger. Both our estimate and that of \citet{brown2016} are of the same order, but considerably smaller than that obtained theoretically by \citet{li2019}, who obtained a density of 1500~kpc$^{-3}$ for ELMs in double degenerate binaries. This shows that there are still inconsistencies between theory and observation that must be solved.

\section{Summary \& Conclusions}
\label{sec:end}

We have compiled a catalogue of known ELMs from multiple sources in the literature, and studied their {\it Gaia} parameters to map the space spanned by ELMs in the observational HR diagram. We selected objects given a colour criterion defined based on the known ELMs ($M < 0.3~M_{\odot}$) with parallax uncertainty smaller than 20 per cent and on evolutionary models, as well as quality control parameters based on \citet{lindegren2018}. Following our selection, we have cross-matched the initial candidate sample with SDSS in order to study those with available spectra. This step revealed a large contamination by main sequence stars and quasars, which we were able to considerably reduce by applying a stricter cut on $\verb!parallax_over_error!$. To further reduce contamination by faint red main sequence stars, we have also studied the $G \times (G-G_{RP})$ diagram, and defined a further selection to exclude objects apparently scattered from the main sequence. Finally, to limit the number of CVs and WD+MS in the sample, we have studied multiple colour-colour diagrams to separate objects with red-excess, as well as X-Ray emission. These objects are also of interest as candidate CVs and WD+MS, and are listed in Appendix~\ref{sec:redstars}. After applying all of these selection criteria, our resulting catalogue of ELM candidates has $5672$ objects, listed in Table~\ref{tab:elm_can_cat}.

The presented catalogue has clearly stated selection criteria, summarised in Table~\ref{tab:selection}, which make it a valuable asset for testing binary evolution models. For example, we predict, given the estimated 9 per cent contamination, a density of 275~kpc$^{-3}$ for ELMs, which is considerably smaller than the theoretical prediction of \citet{li2019}. This might be justified by the incompleteness of the {\it Gaia} DR2 catalogue, which is hard to parse, but suggests we need to further improve the agreement between observations and models. Future {\it Gaia} releases should improve parallax measurements and reduce the necessity of strict quality cuts as those applied here.

\begin{table}
	\centering
	\caption{All the target selection cuts applied to obtain our sample of ELM candidates. $F_{BP}$ and $F_{RP}$ are the {\it Gaia} DR2 fluxes in the $G_{BP}$ and $G_{RP}$ filters, and $\sigma_{BP}$ and $\sigma_{RP}$ are their respective uncertainties.}
	\label{tab:selection}
	\begin{tabular}{c}
	\hline
	\hline
	Selection on {\it Gaia} data\\
	\hline
	$\verb!parallax_over_error! > 5 $\\
	$F_{BP}/\sigma_{BP} > 10 $\\
	$F_{RP}/\sigma_{RP} > 10 $\\
	$E > 1.0 + 0.015\,(G_{BP}-G_{RP}+0.61)^2$\\
	$E < 1.45 + 0.06\, (G_{BP}-G_{RP}+0.61)^2$\\
	$u < 1.2\, \verb!max!(1, \exp(-0.2(G - 19.5)))$\\
	\hline
	\hline
	Selection in the {\it Gaia} observational HR diagram\\
	\hline
	$G_{abs} < 5.25 + 6.94\,(G_{BP}-G_{RP}+0.61)^{1/2.32}$\\
    $G_{abs} > 1.15 (G_{BP}-G_{RP}) + 6.0$\\
	$G_{abs} > -42.2(G_{BP}-G_{RP})^2 + 83.8(G_{BP}-G_{RP}) - 20.1$\\
	\hline
	\hline
    Further colour selections\\
	\hline
	$W_1 - W_2 < 0.0375$\\
	$(g-r) > 1.5\,(r-i) - 0.1$\\
    $(r-i) > 1.8\,(i-z) - 0.1$\\
    $(i-z) < -1.3\,(z-y) + 0.2$\\
	\hline
	\hline
	\end{tabular}
\end{table}

The position in the HR diagram alone does not confirm the nature of our candidates, because of overlap with CVs and WD+MS, as well as contamination by stars scattered from the main sequence due to statistical uncertainties \citep[e.g.][]{pelisoli2019}. Spectra need to be obtained for this sample in order to (i) confirm their nature as ELMs, and (ii) estimate their physical parameters, in order to test predictions from theoretical models. The large multi-object spectroscopic facilities planned for the next years \citep[e.g., WEAVE, 4MOST, DESI, SDSS-V,][]{weave,4most,desi,sdssV} should play a key role in this effort. 

Confirming the nature of these stars will allow us to improve existing models and better understand the evolution of binary stars \citep[e.g.][]{toonen2017}. In particular, objects near the border of $M = 0.3~M_{\odot}$ can shed light on the apparent discrepancy between the mass of evolutionary models and that calculated from spectral models, as well as and help better constrain the lower mass limit for single evolution, which is still a matter of discussion \citep[e.g.][]{kilic2007,justham2010}. Objects in the instability strip \citep[e.g.][]{calcaferro2017} can be used to probe the stellar interiors and possibly rotation rates.

This presented catalogue represents the first step towards a complete and unbiased all-sky catalogue of ELMs. Our completeness is yet limited by the uncertainties in $\it Gaia$ DR2 and should not extend beyond 500~pc, excluding dense regions, but we provide for the first time an all-sky sample extending to temperatures down to 5000~K. This database will be a critically important benchmark in testing models of interacting binaries, with implications also to future gravitational wave experiments.

\section*{Acknowledgements}

IP acknowledges funding by the Deutsche Forschungsgemeinschaft under grant GE2506/12-1 and thanks Stephan Geier, S. O. Kepler, Alina Istrate, and Stephen Justham for useful discussions. This work was supported by a fellowship for postdoctoral researchers from the Alexander von Humboldt Foundation awarded to JV. We thank the anonymous referee for their suggestions, which helped improve our manuscript.

This work has made use of data from the European Space Agency (ESA) mission Gaia (\url{https://www.cosmos.esa.int/gaia}), processed by the Gaia Data Processing and Analysis Consortium (DPAC, \url{https://www.cosmos.esa.int/web/gaia/dpac/consortium}). Funding for the DPAC has been provided by national institutions, in particular the institutions participating in the Gaia Multilateral Agreement. This research has also made extensive use of TOPCAT \citep[\url{http://www.starlink.ac.uk/topcat/},][]{taylor2005}, and of NASA's Astrophysics Data System (\url{http://adsabs.harvard.edu/}).




\bibliographystyle{mnras}
\bibliography{ELMcat} 




\appendix

\section{List of ELM companions to milisecond pulsars}
\label{sec:apA}

\begin{table*}
	\centering
	\caption{ELM stars as companions to milisecond pulsars. This updates table~1 of \citet{smedley2014}, without including tentative identifications such as objects in \citet{mignani2014}.}
	\label{tab:PSRs}
	\begin{tabular}{cccccccccc}
	\hline
	\hline
	PSR & RA & DEC & $M_\textrm{WD} $ & $\sigma_\textrm{WD} $ & $M_\textrm{PSR}$ & $\sigma_{M_\textrm{PSR}} $ & P$_\textrm{orb}$ & Reference \\
	  &  (J2000)  &  (J2000) & $(M_{\odot})$ & $(M_{\odot})$ & $(M_{\odot})$ & $(M_{\odot})$ & (days) & \\ 
	\hline
	J0218+4232 & 02:18:06.36 & +42:32:17.4 & 0.2 &  & 1.6 &  & 2.029 & \citet{bassa2003} \\
	J0348+0432 & 03:48:43.64 & +04:32:11.5 & 0.172 & 0.003 & 2.01 & 0.04 & 0.1025 & \citet{antoniadis2013} \\
	J0437-4715 & 04:37:15.93 & -47:15:09.3 & 0.254 & 0.014 & 1.58 & 0.18 & 5.74 & \citet{2018CQGra..35m3001V}\\
	J0614-3329 & 06:14:10.35 & -33:29:54.1 & 0.24 & 0.04 &  &  & 53.585 &  \citet{2016MNRAS.455.3806B}\\
	J0751+1807 & 07:51:09.15 & +18:07:38.3 & 0.191 & 0.015 & 1.26 & 0.14 & 0.26 & \citet{2008AIPC..983..453N}\\
	J1012+5307 & 10:12:33.43 & +53:07:02.6 & 0.156 & 0.02 & 1.64 & 0.22 & 0.6047 & \citet{1998MNRAS.298..207C}\\
	J1713+0747 & 17:13:49.53 & +07:47:37.5 & 0.28 & 0.03 & 1.3 & 0.2 & 67.83 & \citet{2005ApJ...620..405S}\\
	J1738+0333 & 17:38:53.96 & +03:33:10.8 & 0.181 & 0.006 & 0.181 & 0.006 & 0.35 & \citet{2015MNRAS.446L..26K}\\
	J1853+1303 & 18:53:57.32 & +13:03:44.1 & 0.35 & 0.2 & 1.4 & 0.7 & 115.65 & \citet{2011ApJ...743..102G}\\
	J1857+0943 & 18:57:36.39 & +09:43:17.2 & 0.27 & 0.025 & 1.6 & 0.2 & 12.327 & \citet{2004PhDT........21S}\\
	J1909-3744 & 19:09:47.44 & -37:44:14.3 & 0.204 & 0.002 & 1.438 & 0.024 & 1.53 & \citet{2015MNRAS.446L..26K}\\
	J1910+1256 & 19:10:09.70 & +12:56:25.5 & 0.315 & 0.015 & 1.6 & 0.6 & 58.47 & \citet{2011ApJ...743..102G}\\
	J1911-5958A & 19:07:50.53 & +12:51:28.3 & 0.18 & 0.02 & 1.4 &  & 0.865 & \citet{bassa2006}\\
	J1959+2048 & 19:59:36.75 & +20:48:14.6 & 0.035 & 0.002 & 2.4 &  & 0.382 & \citet{2011ApJ...743..102G}\\
	J2016+1948 & 20:16:56.70 & +19:48:03.0 & 0.45 & 0.02 & 1.0 & 0.5 & 635.04 & \citet{2011ApJ...743..102G}\\
	J2317+1439 & 23:17:09.23 & +14:39:31.2 & 0.39 & 0.13 & 3.4 & 1.4 & 2.459 & \citet{2017ApJ...842..105D}\\
	\hline
	\hline
	\end{tabular}
\end{table*}

\onecolumn{

\section{ADQL queries}

\subsection{Inital query}
\label{ref:initial}

\begin{verbatim}
SELECT source_id, ra, dec, parallax, parallax_error, pmra, pmra_error, pmdec,
       pmdec_error, phot_g_mean_mag, phot_bp_mean_mag,
       phot_rp_mean_mag, bp_rp, parallax_over_error, phot_g_mean_flux_over_error,
       phot_bp_mean_flux_over_error, phot_rp_mean_flux_over_error,
       phot_bp_rp_excess_factor, astrometric_excess_noise,
       astrometric_chi2_al, astrometric_n_good_obs_al
FROM gaiadr2.gaia_source
WHERE parallax_over_error > 3
AND phot_bp_mean_flux_over_error>10
AND phot_rp_mean_flux_over_error>10
AND phot_bp_rp_excess_factor < 1.45+0.06*power(phot_bp_mean_mag-phot_rp_mean_mag,2)
AND phot_bp_rp_excess_factor > 1.0+0.015*power(phot_bp_mean_mag-phot_rp_mean_mag,2)
AND ( astrometric_chi2_al/(astrometric_n_good_obs_al-5)<1.44
     OR astrometric_chi2_al/(astrometric_n_good_obs_al-5)<1.44*exp(-0.4*(phot_g_mean_mag-19.5)) )
AND ( bp_rp < 1.03733 AND bp_rp > -0.60999 AND
     5+5*log10((parallax+0.029)/1000)+phot_g_mean_mag < 5.25 + 6.94*power((bp_rp+0.61),(1./2.32)) )
AND ( (bp_rp > 0.688482 AND bp_rp < 1.03733
     AND 5+5*log10((parallax+0.029)/1000)+phot_g_mean_mag > 20.25*bp_rp-7.15 )
     OR (bp_rp > -0.60999 AND bp_rp < 0.688482 AND
     5+5*log10((parallax+0.029)/1000)+phot_g_mean_mag > 1.15*bp_rp + 6.00 ) )
\end{verbatim}


\subsection{Refined query}
\label{sec:refined}

\begin{verbatim}
SELECT source_id, ra, dec, parallax, parallax_error, pmra, pmra_error, pmdec,
       pmdec_error, phot_g_mean_mag, phot_bp_mean_mag,
       phot_rp_mean_mag, bp_rp, parallax_over_error, phot_g_mean_flux_over_error,
       phot_bp_mean_flux_over_error, phot_rp_mean_flux_over_error,
       phot_bp_rp_excess_factor, astrometric_excess_noise,
       astrometric_chi2_al, astrometric_n_good_obs_al
FROM gaiadr2.gaia_source
WHERE parallax_over_error > 5
AND phot_bp_mean_flux_over_error>10
AND phot_rp_mean_flux_over_error>10
AND phot_bp_rp_excess_factor < 1.45+0.06*power(phot_bp_mean_mag-phot_rp_mean_mag,2)
AND phot_bp_rp_excess_factor > 1.0+0.015*power(phot_bp_mean_mag-phot_rp_mean_mag,2)
AND ( astrometric_chi2_al/(astrometric_n_good_obs_al-5)<1.44
    OR astrometric_chi2_al/(astrometric_n_good_obs_al-5)<1.44*exp(-0.4*(phot_g_mean_mag-19.5)) )
AND ( bp_rp < 0.7803 AND bp_rp > -0.60999 AND
     5+5*log10((parallax+0.029)/1000)+phot_g_mean_mag < 5.25 + 6.94*power((bp_rp+0.61),(1./2.32)) )
AND ( (bp_rp > 0.5808 AND bp_rp < 0.7803 AND
     5+5*log10((parallax+0.029)/1000)+phot_g_mean_mag > 33*bp_rp-12.5 ) OR
     (bp_rp > -0.60999 AND bp_rp < 0.5808 AND
     5+5*log10((parallax+0.029)/1000)+phot_g_mean_mag > 1.15*bp_rp + 6.00 ) )
\end{verbatim}

}

\section{CV and WD+M candidates}
\label{sec:redstars}

\begin{table}
	\centering
	\caption{Objects that were separated from our ELM candidate selection due to X-ray detection or red-excess. They should be mostly CVs and white dwarfs with main sequence companions. The full table with 3526 objects can be found in the online version of the paper.}
	\label{tab:cvdam}
	\begin{tabular}{cccccccccccc}
	\hline
	\hline
	$\verb!source_id!$ & RA & DEC & Parallax & $\mu$ & G \\
  	 & (J2000) & (J2000) & (mas) & (mas/yr) & (mag) \\
  	 \hline
384877723112816128 & 00:00:27.76 & +43:32:47.1 & 1.550 & 15.793 & 17.857\\
2846765106166697472 & 00:01:02.14 & +21:34:48.1 & 4.330 & 28.879 & 19.375\\
4635348028348773120 & 00:01:14.58 & -79:29:52.7 & 1.915 & 8.436 & 19.496\\
2880918381162658560 & 00:01:45.04 & +38:13:05.4 & 3.676 & 2.460 & 19.777\\
2307584090171983616 & 00:02:07.36 & -37:49:16.7 & 2.178 & 53.612 & 17.115\\
2873571577609388288 & 00:03:17.21 & +31:51:38.9 & 4.126 & 39.972 & 18.977\\
2873491347619449216 & 00:03:29.81 & +31:18:12.0 & 4.145 & 10.811 & 19.337\\
2874194833198918400 & 00:03:42.15 & +32:44:15.9 & 3.208 & 12.608 & 18.246\\
2873505301968425984 & 00:03:46.01 & +31:37:29.9 & 2.429 & 28.233 & 18.006\\
574031347003008768 & 00:04:25.32 & +85:52:53.2 & 2.166 & 14.980 & 19.885\\

	\hline
	\hline
	\end{tabular}
\end{table}


\bsp	
\label{lastpage}
\end{document}